# Free-form Lesion Synthesis Using a Partial Convolution Generative Adversarial Network for Enhanced Deep Learning Liver Tumor Segmentation

Yingao Liu[1], Fei Yang[2], Yidong Yang [3,1*]

[1] Department of Engineering and Applied Physics, University of Science and Technology of China, Hefei, Anhui, 230026 China

[2] Department of Radiation Oncology, University of Miami, Florida, USA

[3] Department of Radiation Oncology, the First Affiliated Hospital of USTC. Division of Life Sciences and Medicine, University of Science and Technology of China, Hefei, Anhui, 230026 China

*Corresponding author:

Yidong Yang, PhD, DABR

Professor and Director of Medical Physics Program, Department of Engineering and Applied Physics，School of Physical Sciences, University of Science and Technology of China

E-mail: ydyang@ustc.edu.cn

**Abstract:**

Lesion segmentation is critical for clinicians to accurately stage the disease and determine treatment strategy. Automatic deep learning segmentation models has been shown to improve both the segmentation efficiency and the accuracy. However, training a robust segmentation model requires considerably large labeled training samples, which may be impractical. This study aimed to develop a deep learning framework for generating synthetic lesions that can be used to enhance network training. The lesion synthesis network is a modified generative adversarial network (GAN). Specifically, we innovated a partial convolution strategy to construct an Unet-like generator. The discriminator is designed using Wasserstein GAN with gradient penalty and spectral normalization. A mask generation method based on principal component analysis was developed to model various lesion shapes. The generated masks are then converted into liver lesions through a lesion synthesis network. The lesion synthesis framework was evaluated for lesion textures, and the synthetic lesions were used to train a lesion segmentation network to further validate the effectiveness of this framework. All the networks are trained and tested on the public dataset from LITS. The synthetic lesions generated by the proposed approach have very similar histogram distributions compared to the real lesions for the two employed texture parameters, GLCM-energy and GLCM-correlation. The Kullback-Leibler divergence of GLCM-energy and GLCM-correlation were 0.01 and 0.10, respectively. Including the synthetic lesions in the tumor segmentation network improved the segmentation dice performance of U-Net significantly from 67.3% to 71.4% ($p<0.05$). Meanwhile, the volume precision and sensitivity improve from 74.6% to 76.0% ($p=0.23$) and 66.1% to 70.9% ($p<0.01$), respectively. The proposed lesion synthesis approach can be used for free-form lesion generation to produce additional labeled training samples. The synthetic data significantly improves the segmentation performance. The approach shows great potential for alleviating the "data paucity" problem.

# 1. INTRODUCTION

In recent years, deep learning has made significant achievements in medical image segmentation(Wu *et al* 2020),(Saood and Hatem 2021). However, a large amount of labeled data covering sufficient data diversity is necessary for the development of a robust deep learning model. Manual annotation of medical images is a time-consuming and labor-intensive task that can only be accomplished by experienced clinical specialists, so it is difficult to collect a sufficiently large amount of labeled data. On the other hand, the collection of medical images often requires a dedicated protocol to obtain consent, and hence is a non-trivial process. Therefore, how to generate labeled data more effectively and efficiently is a critical challenge. One way to automatically generate labeled data is to synthesize lesions using a deep learning network. The state-of-art for lesion synthesis uses generative adversarial networks (Ian J. Goodfellow, Jean Pouget-Abadie, Mehdi Mirza, Bing Xu, David Warde-Farley, Sherjil Ozair, Aaron Courville 2017) (GANs). Frid-Adar et al. (Frid-Adar *et al* 2018) utilized a GAN to generate synthetic images to improve liver lesion classification. The initial GAN usually generates images without any labeling. In the segmentation, however, images with lesion contours are necessary. Isola et al. (Isola *et al* 2017) introduced conditional GAN (cGAN) which can generate new images with classified data labels. Abhishek et al. (Abhishek and Hamarneh 2019) applied the method to synthesize lesions while keeping their original contours and then used the synthesized lesions to augment the training dataset for enhanced lesion segmentation. Likewise, Jin et al. (D. Jin, Z. Xu, Y. Tang, A.P. Harrison 2018) further developed a 3-dimensional (3D) cGAN to simulate labeled lung nodules for enhanced lung lesion segmentation.

Despite the promising achievements of GAN, the standard convolution operation in GAN applies the same filters to all image pixels, both inside and outside the mask, inevitably leading to blurred lesion borders or lesion texture loss. To deal with this problem, partial convolution(Liu, G., Reda, F.A., Shih, K.J., Wang, T.-C., Tao, A., Catanzaro 2018) is proposed to apply filters only to pixels outside a mask. Dong et al.(Dong *et al* 2021) utilized 3D partial convolution to reconstruct the missing regions in ultrasound images using least-squares generative adversarial network(Mao *et al* 2017) (LSGAN) network. Zhang et al.(Zhang *et al*

2021) employed partial convolution to generate synthetic hemorrhage lesions for improved intracranial hemorrhage diagnosis using the cGAN network.

Inspired by these methods, we develop a new partial convolution GAN (PCGAN) to generate synthetic lesions with predefined lesion contours and realistic textures. To penalize the content and texture losses of synthetic lesions, we define a hybrid loss function, which includes style, content, and reconstruction losses. Meanwhile, the adversarial loss in Wasserstein GAN (WGAN) is added to the loss function to further enhance the synthesis results. Thereafter, to stabilize the training process, the GAN network is optimized by adding spectral normalization and gradient penalty to the discriminator. The proposed network is used to generate additional labeled training images to enhance the performance of deep learning-based lesion segmentation.

## 2. MATERIALS AND METHODS

### 2.1 Image dataset

In this study, we used the public dataset from LITS(Bilic *et al* 2019) which contains 131 subjects with liver cancer. We selected lesions larger than 10 pixels for reliable lesion texture assessment(Pan *et al* 2021), resulting in 6612 images with lesions from 117 subjects. To accelerate the training process, all the CT images were resized to 256×256 pixels from 512×512 pixels. The liver in each image was extracted using the contour drawn by physicians. The image window was set to [-100, 200] HU, and the intensity normalized to [0,1]. In our work, we use two kinds of networks, a lesion texture generation network to generate synthetic lesions and a segmentation network for lesion segmentation. All images were divided into three sets such that the training/ validation/testing set each contains 3454/1618/1540 images from 87/18/12 patient CT scans. While the data division was same, the training strategies were different. The lesion texture generation network did not use cross-validation, while the segmentation network was trained in five-fold cross-validation to obtain more reliable segmentation results.

### 2.2 Liver lesion synthesis

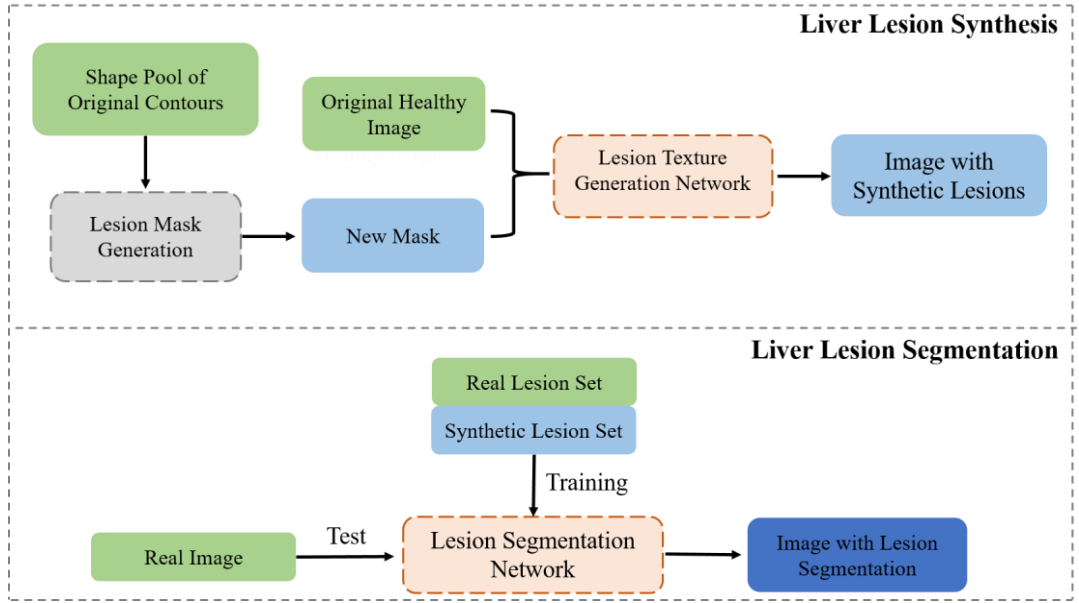

Figure 1: The workflow of our approach.

Figure 1 depicts the workflow of the proposed lesion segmentation approach. Liver lesions appear in various shapes. The shape of a lesion is an important parameter and should be modeled reasonably by mimicking real ones. The proposed lesion segmentation strategy consists of three stages. The first stage is to automatically generate masks from a shape pool containing all the manual contours from the real lesions in the training dataset. The second stage is to produce synthetic lesions on CT images bearing no lesions by applying a novel lesion texture synthesis network. The last stage is to segment real lesions using a deep learning network that is trained with both real and synthetic lesions.

**2.2.1 Lesion mask generation**

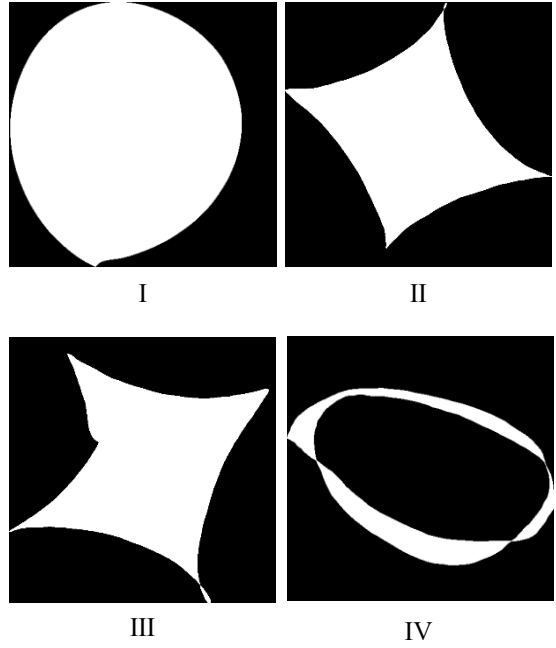

Figure 2: 1$^{st}$, 2$^{nd}$, 3$^{rd}$ and 4$^{th}$ principal mode extracted from the PCA modeling of existing lesion shapes.

First, we used principal component analysis (PCA) to perform shape decomposition and to analyze the shape variation. All the lesion masks are aligned with each other and then scaled to the same area. Empirically, 200 points equally spaced along the boundary of a mask were extracted to form a vector $[x_1, x_2, ..., x_n, y_1, y_2, ..., y_n]^T$ representing the shape of the mask, where $x_i$ and $y_i$ are the position of the i$^{th}$ point. Then the Procrustes analysis(Goodall 1991) is employed to mathematically register all the shapes by taking into account the rotation and scaling effects . Finally, PCA is employed to model the shape variation in the mask pool. Ten principal components are used, and the first four principal components are demonstrated in Figure 2. The first component can be interpreted as the shape average of all the masks, while the remaining modes reflect the shape variation at different orders of magnitude. We use $D = [D_1, D_2, ..., D_{10}]^T$ to denote the first 10 decomposed components and $w = [w_1, w_2, ..., w_{10}]$ the corresponding weight. Thus, a new shape can be generated through a transformation operation $T(wD)$ by varying the weight w. Here the purpose of the transformation $T(\cdot)$ is to diversify the location, size, and rotation of the synthetic lesion.

### 2.2.2 Lesion texture generation

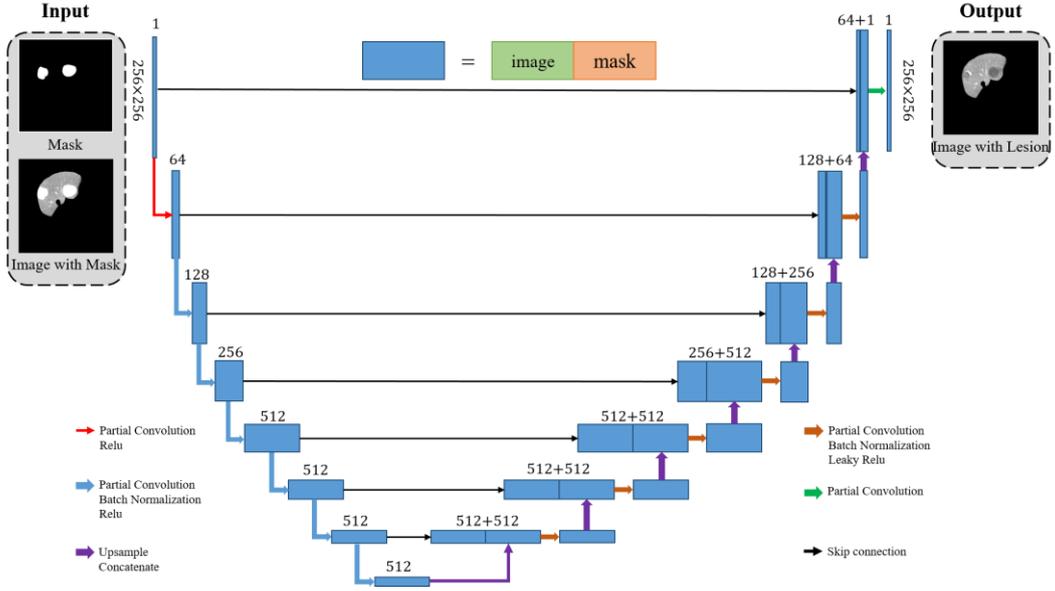

Figure 3: The structure of the U-Net lesion texture generator. It is built with partial convolutional layers. The input is a binary mask and a masked image. The output is an image with synthetic lesions. The encoder and decoder part each has eight stages. Skip connection is applied to each stage.

The generator, illustrated in Figure 3, is built with partial convolutional layers. The input is a binary mask and a CT image bearing this mask. In the training stage, the mask is obtained from physicians' lesion delineation. In the testing stage, automatically generated masks are applied. The purpose of using partial convolution is to make the convolution process dependent only on the unmasked pixels via a re-normalization step. The partial convolution is defined as follows:

$$X'_{(i,j)} = \begin{cases} W^T(X_{(i,j)} \odot M_{(i,j)}) \frac{1}{M_{(i,j)}} + b, & if\ sum(M_{(i,j)}) > 0 \\ 0, & otherwise \end{cases} \quad (1)$$

where $X'_{(i,j)}$ is the convolution result at position $(i,j)$. W is the convolutional filter weight and b is the bias of the filter. X is the feature value for the convolution window at the position $(i,j)$ and M is the corresponding binary mask (1 for unmasked pixels, 0 for masked pixels). X is updated from layer to layer.

Meanwhile, the binary mask is also simultaneously updated with the convolution process via a customized operation. In the operation, a kernel with size same as that of the convolution window is first defined. If there is any pixel in the mask that has a value of 1, then all the pixels covered by the kernel will be assigned the value 1. After a sufficient number of updates, all the pixels in the mask will finally have a value of 1, and all the pixels in the image will become unmasked.

The generator is a Unet(Ronneberger *et al* 2015) variant and consists of 8 stages. ReLu is used as the activation function in the encoder part, while Leaky ReLu with a slope parameter of 0.2 is used in the decoder part. Batch normalization(Ioffe and Szegedy 2015) is adopted for all the convolutional layers except the first and last layers. Skip connection is also introduced to ensure feature reusability by concatenating the feature maps of the encoders and decoders of the same stages.

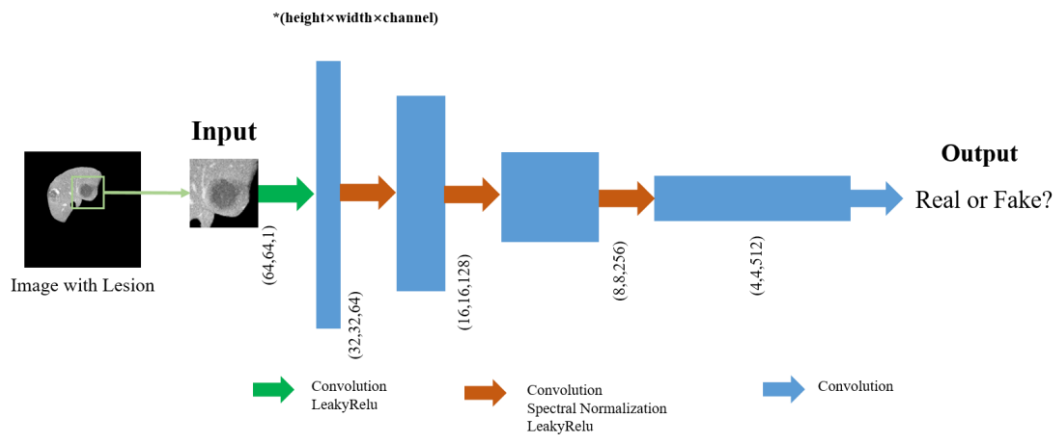

Figure 4: The structure of the discriminator. The input is a 64x64 matrix with a randomly-selected single lesion. There are 4 layers in the discriminator. Each of the first three layers has a kernel size of 4, a stride of 2, and a padding of 1. In the last layer, the output is reduced to one channel and obtain a prediction probability.

The architecture of the discriminator $D_l$ is illustrated in Figure 4. The input is a 64x64 matrix with a randomly-selected single lesion. There are 4 layers in the discriminator. Each of the first three layers has a kernel size of 4, a stride of 2, and a padding of 1. In the last layer, the output is reduced to one channel and provides a prediction probability. Spectral normalization(Masa, Takeru Miyato 2018) is employed after all intermediate convolutional layers to stabilize the training of the networks.

The generator and the discriminator are trained to minimize a total loss function, which is defined as:

$$L_{total} = L_{GAN} + \lambda_{reconstruction}L_{reconstruction} + \lambda_{perceptual}L_{perceptual} + \lambda_{texture}L_{texture} \quad (2)$$

$\lambda_{reconstruction}$, $\lambda_{perceptual}$ and $\lambda_{texture}$ are introduced mainly to weigh the importance of the four losses. After optimization, we set $\lambda_{reconstruction} = 1$, $\lambda_{perceptual} = 0.05$ and $\lambda_{texture} = 100$.

$L_{GAN}$ is designed to encourage the generator to generate more similar textures between the generated and real lesions with the supervision of the discriminator. The Wasserstein GAN(Gulrajani *et al* 2017) (WGAN) loss is chosen as our loss function. The generator and discriminator are trained by solving an adversarial problem: $\min_G \max_D L_{GAN}(G, D)$, and the GAN loss is defined as:

$$L_{GAN}(G, D) = E[D(x_i \odot M_i)] - E[D(\hat{x}_i \odot M_i)] - \lambda_{gp} G_p(D) \quad (3)$$

where $x_i$, $\hat{x}_i$, and $M_i$ represent the $i^{th}$ real image, its corresponding synthetic image, and binary mask, respectively. $G_p(D)$ is the gradient penalty to enforce the Lipschitz constraint.

$L_{reconstruction}$ is designed to calculate the difference between the ground truth and the predicted image, and is defined as:

$$L_{reconstruction} = w_1 \|M_i \odot (x_i - \hat{x}_i)\|_1 + w_2 \|(1 - M_i) \odot (x_i - \hat{x}_i)\|_1 \quad (4)$$

where $\odot$ denotes element-wise multiplication. $w_1$, $w_2$ are used to weigh the importance of healthy tissue and lesion. After optimization, we set $w_1 = 1$ and $w_2 = 5$.

$L_{perceptual}$ is utilized to capture the high-level semantics and simulate the human perception of image quality. Perceptual loss(Johnson *et al* 2016) is defined as:

$$L_{perceptual} = E[\sum_i \frac{1}{N_i} (\|\Phi_i(\hat{x}_i) - \Phi_i(x_i)\|_1 + \|\Phi_i(z) - \Phi_i(x_i)\|_1)] \quad (5)$$

where $N_i$ denotes the number of pixels in $x_i$, $\Phi_i$ is the feature map of the $i^{th}$ layer of VGG-16(Simonyan, K., Zisserman 2014) network pre-trained on ImageNet(Deng, J., Dong, W., Socher, R., Li, L.-J., Li, K., Fei-Fei 2009). z is a composite of the real image and the predicted image, which is defined as $z = M_i \odot x_i + (1 - M_i) \odot \hat{x}_i$.

$L_{texture}$ has the same form as $L_{perceptual}$. But the former does not act directly on the feature maps as the latter does. Instead, L is calculated based on the autocorrelation (Gram

matrix)(Gatys, L., Ecker, A.S., Bethge 2015) of all the feature maps. It doesn't consider the individual pixel position but focuses on the correlation between pixels. Given any feature map of size $C_j \times H_j \times W_j$, the texture loss is defined as follows:

$$L_{texture} = E_j[\sum_i \frac{1}{N_i}(\|G_j^{\Phi}(\hat{x}_i) - G_j^{\Phi}(x_i)\|_1 + \|G_j^{\Phi}(z) - G_j^{\Phi}(x_i)\|_1)] \qquad (6)$$

where $G_j^{\Phi}$ is a $C_j \times C_j$ Gram matrix constructed from the selected feature maps.

The lesion synthesis network is optimized using AMSGrad(Reddi *et al* 2019) with decay factor for the first moment ($\beta_1$) of 0.5 and decay factor for infinity norm ($\beta_2$) of 0.999and is trained for 500000 iterations. Kernels are initialized using the method described by He et al (He *et al* 2015). The generator and discriminator are simultaneously updated, and the learning rates are set to 0.0001 and 0.00001, respectively. The training was conducted on an NVIDIA RTX 3090 GPU (24GB) with a batch size of 6 and took 28 hours.

### 2.2.3 Evaluation of synthetic lesions

Radiomics features including gray-level cooccurrence matrix-based energy and correlation are utilized to evaluate the realism of the synthetic lesions. We computed the feature distributions of synthetic lesions and compared them with real lesions. The similarity between the distributions was assessed by using Kullback-Leibler (KL) divergence, which is given as follows:

$$KL(h_1, h_2) = \sum_{i=0}^{n-1} h_1(i) \ln \frac{h_1(i)}{h_2(i)} \qquad (7)$$

where n is the bins of the histogram. $h_1$ and $h_2$ are normalized histograms. $h(i)$ denotes the height of the i[th] bin. $KL$ is to calculate the asymmetry of the difference between two discrete distributions $h_1$ and $h_2$.

The synthetic lesions are generated using the proposed PCGAN, and the results are benchmarked against two other typical methods, including cGAN and Tub-sGAN. As for cGAN, the network architecture reported by Abhishek(Abhishek and Hamarneh 2019) et al was used which performs well in skin lesion synthesis for enhanced lesion segmentation. The Tub-sGAN(Zhao *et al* 2018) network is the first to incorporate style transfer into the GAN framework. It adopted a cGAN architecture that combined style, content, and L1 losses.

## 2.3 Enhanced lesion segmentation

### 2.3.1 Network architecture

Three widely used semantic segmentation networks: U-Net, Attention Unet(Oktay *et al* 2018), and Unet++(Zhou *et al* 2018) are employed. We adopted the original architecture of these networks. These three networks have the same batch size, epoch, and learning rate in the training process. The batch size is set to 16, the epoch to 150, and the learning rate to 0.0003, respectively. A combination of cross-entropy loss(Yi-de *et al* 2004) and Dice loss(Sudre *et al* 2017) is applied as the loss function, which is defined as:

$$L_{CE-Dice} = \lambda_{CE} L_{CE} + \lambda_{Dice} L_{Dice} \tag{8}$$

With $L_{CE}(y, \hat{y}) = -\frac{1}{N}\sum_i y \log(\hat{y}) + (1-y)\log(1-\hat{y})$, $L_{Dice}(y, \hat{y}) = 1 - \frac{2y\hat{y}+1}{y+\hat{y}+1}$

where $\hat{y}$ refers to the predicted value and $y$ to the ground truth label. In Dice loss, 1 is added in both numerator and denominator to ensure that the function is not undefined in boundary scenarios such as when $y = \hat{y} = 0$. $\lambda_{Dice}$ and $\lambda_{CE}$ are set to 1 and 0.5, respectively.

### 2.3.2 Evaluation of segmentation

The metrics used to evaluate the segmentation performance include dice similarity coefficient (DSC), volume precision (vPSC), and volume sensitivity(vSEN) such as:

$$\text{DSC}(\%) = \frac{2(V_{pre} \cap V_{gt})}{V_{pre} + V_{gt}} \times 100\% \tag{9}$$

$$\text{vPSC}(\%) = \frac{V_{pre} \cap V_{gt}}{V_{pre}} \times 100\% \tag{10}$$

$$\text{vSEN}(\%) = \frac{V_{pre} \cap V_{gt}}{V_{gt}} \times 100\% \tag{11}$$

where $V_{pre}$ and $V_{gt}$ represent the predicted and ground truth lesion volume.

## 3. RESULTS

### 3.1. Evaluation of synthetic lesions

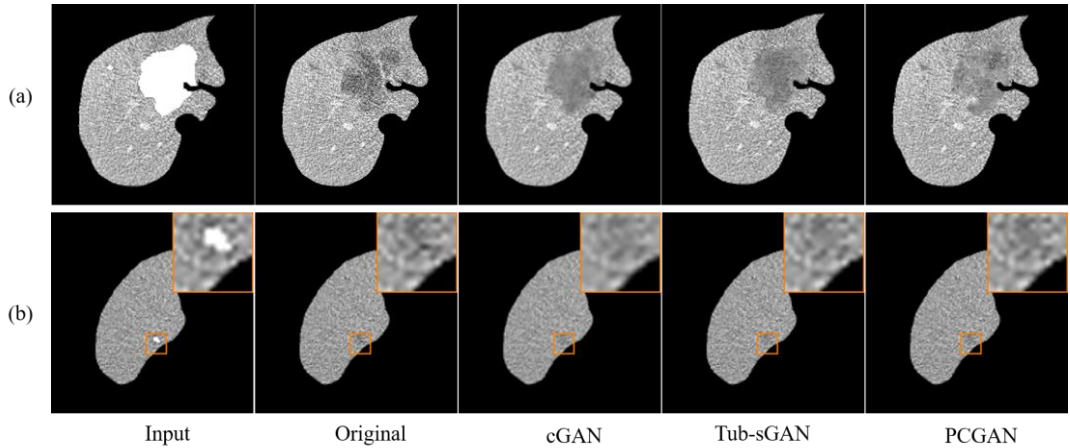

Figure 5: Synthesis results of different lesion synthesis methods using two infected slices and their corresponding manual contoured masks in the test set.

The similarity in the image appearance between the synthetic and real lesions was first evaluated. Figure 5 compares the synthetic lesions generated on real infected image slices using different texture-generation networks. The original lesion contour manually delineated by physicians was used as the lesion mask. Figure 5(a) shows the lesion generated by the proposed network PCGAN blend well into the surrounding healthy tissues and has a more heterogeneous texture than those generated with cGAN and Tub-sGAN. Figure 5(b) gives an example of a small lesion about 2-mm in dimension. The lesion boundary is almost lost for cGAN, but is well maintained for the PCGAN.

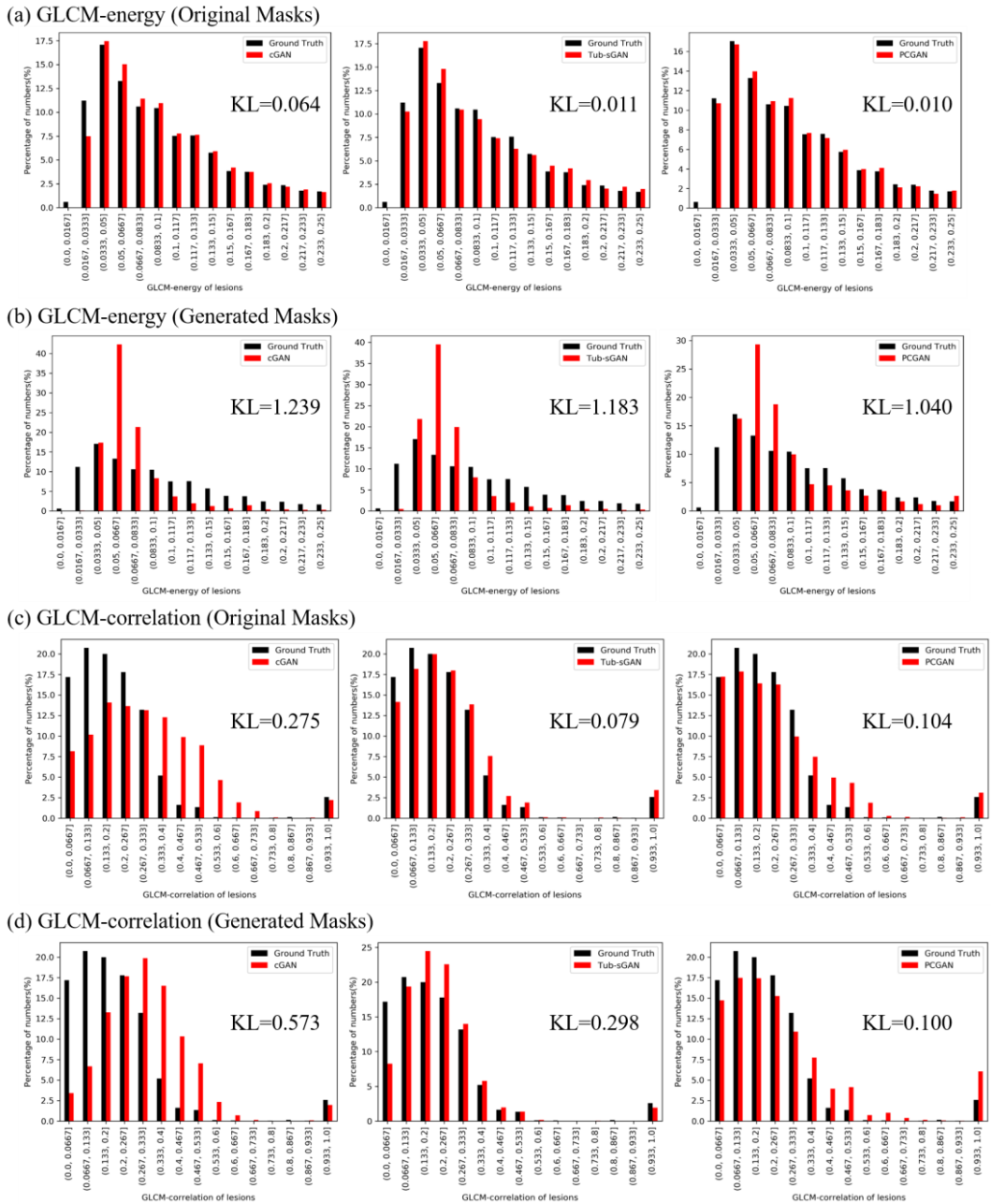

Figure 6: The radiomics features histograms for real images and synthetic images generated by different lesion synthesis methods. Kullback-Leibler (KL) divergence is applied to compute the distances of two radiomics features distributions.

Two commonly used radiomics features, GLCM-energy and GLCM-correlation, were applied to compare the collective similarity between the synthetic and real lesion sets. As shown in Figure 6, the synthetic and real lesion sets have very similar distributions for the two parameters GLCM-energy and GLCM-correlation, particularly for those generated using the original contours. For GLCM-energy, PCGAN has the KL score of 0.010 and performs best in these three methods. For GLCM-correlation, Tub-sGAN and PCGAN perform better, with their KL scores of 0.079 and 0.104, respectively. The GLCM-correlation distributions for those created with generated masks still achieve low KL scores, and PCGAN has the lowest KL score of 0.100. However, the GLCM-energy distributions for those created with generated masks are slightly off from the ground truth. The KL scores for cGAN, Tub-sGAN, and PCGAN are 1.239, 1.183, and 1.040, respectively.

### 3.2. Evaluation of enhanced lesion segmentation

Figure 7 demonstrates the lesion segmentation performance when including the synthetic lesions in the training data set. The lesion segmentation network was a classic U-Net that is heavily used in segmentation tasks. The demonstrated three lesions vary in location, size, and shape. Interestingly, Figure 7(a) gives an extreme case with 15 separate lesions. The proposed network PCGAN produced both fewer false positives and fewer false negatives. However, it still misses two small lesions, as indicated by the orange arrows in Figure 7(a). In Figure 7 (b) and (c), PCGAN delineates more accurate lesion boundaries. Particularly for the small lesion with an irregular shape, PCGAN segmented almost all the lesion volume while the other two networks missed most of it.

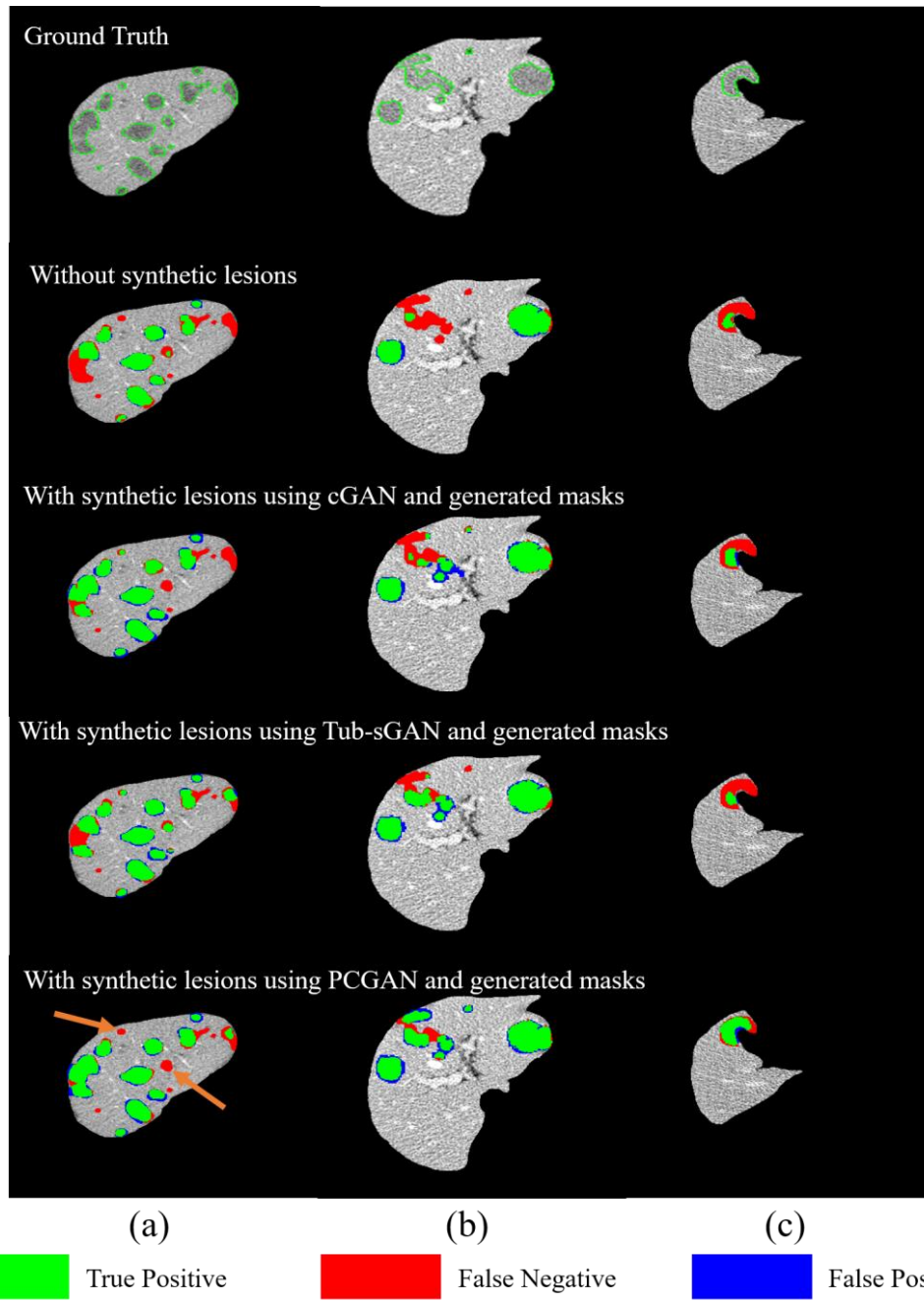

Figure 7: Comparison of enhanced lesion segmentation using networks trained with different lesion synthesis methods. (a–c) show segmentation results from different patients. The green, red, and blue areas are true positive, false negative, and false positive, respectively.

Table 1: Segmentation performance using lesions synthesized with different methods. The best results are highlighted in bold. "Real" and "Syn" refer to 3454 real images and 1731 synthetic images. Both real images and synthetic images are used to train segmentation networks. "Averaged" is the average result.

|  | Methods | DSC(%) | vPSC(%) | vSEN(%) |
|---|---|---|---|---|
|  | Real | 67.3 ± 19.0 | 74.6 ± 21.6 | 66.1 ± 17.0 |
|  | Real+Syn[Averaged] | 70.0 ± 17.2 | 75.7 ± 20.4 | 68.9 ± 15.5 |
| U-Net | Real+Syn[cGAN] | 69.1 ± 18.1 | 74.6 ± 21.5 | 68.9 ± 15.0 |
|  | Real+Syn[Tub-sGAN] | 69.4 ± 17.2 | **76.5 ± 20.0** | 66.8 ± 15.6 |
|  | Real+Syn[PCGAN] | **71.4 ± 16.2** | 76.0 ± 19.5 | **70.9 ± 14.7** |
|  | Real | 64.9 ± 22.0 | 71.0 ± 24.3 | 66.7 ± 17.6 |
|  | Real+Syn[Averaged] | 69.4 ± 18.7 | 73.7 ± 21.8 | 70.0 ± 15.1 |
| Attention U-Net | Real+Syn[cGAN] | 68.6 ± 20.0 | 71.6 ± 23.6 | 70.9 ± 14.7 |
|  | Real+Syn[Tub-sGAN] | 69.2 ± 18.3 | **75.0 ± 21.0** | 68.1 ± 15.9 |
|  | Real+Syn[PCGAN] | **70.3 ± 17.7** | 74.6 ± 20.5 | **71.0 ± 14.6** |
|  | Real | 68.0 ± 18.8 | 75.0 ± 20.8 | 66.8 ± 18.0 |
|  | Real+Syn[Averaged] | 69.8 ± 17.4 | 77.3 ± 20.0 | 67.6 ± 16.8 |
| Unet++ | Real+Syn[cGAN] | 70.1 ± 17.2 | 76.2 ± 20.8 | 69.1 ± 15.2 |
|  | Real+Syn[Tub-sGAN] | 67.1 ± 18.9 | **79.1 ± 20.2** | 62.2 ± 19.1 |
|  | Real+Syn[PCGAN] | **72.1 ± 15.7** | 76.7 ± 18.8 | **71.4 ± 14.2** |

The segmentation performance on the testing dataset is given in Table 1. The segmentation results are obtained by training a U-Net/Attention U-Net/Unet++ using the real training data and synthetic images generated by different lesion synthesis methods. As shown in Table 1, when the training dataset is augmented with the synthetic images generated by different synthesis methods, the segmentation performance improves significantly. Compared with the original segmentation results, the averaged results of these three lesion synthesis methods reach a DSC difference of 2.7% and 4.5% in U-Net and Attention U-Net, and both are significant ($p<0.05$). In Unet++, the DSC difference is 1.8% and is not significant. The differences of vPSC and vSEN are 1.1%, 2.8% in U-Net, 2.7%, 3.3% in Attention U-Net and 2.3%, 0.8% in Unet++.

Compared with other lesion synthesis methods, the proposed PCGAN results in higher segmentation scores in DSC and vSEN. The differences in DSC are 4.1% in U-Net, 5.4% in Attention U-Net, and 4.1% in Unet++, and are all significant ($p<0.05$). The differences of vPSC and vSEN are 1.4%, 4.8% in U-Net, 3.6%, 4.3% in Attention U-Net and 1.7%, 4.6% in Unet++.

## 4. DISCUSSION

In this study, we propose a lesion synthesis method to generate labeled training image samples for enhanced lesion segmentation. When the training dataset is augmented with the synthetic images, all the segmentation networks under evaluation achieved improved performance, due to the increased lesion diversity in the training dataset brought in by the synthetic lesions. Compared with other lesion synthesis methods, cGAN and Tub-sGAN, the proposed PCGAN can generate lesions with more realistic textures.

There are still some limitations of the proposed method. First, the texture evaluation needs further investigation. The current work focuses on using GLCM features for texture evaluation because it represents pixel-wise spatial relationship well. Previously, Pan et al.(Pan *et al* 2021) evaluated the synthetic lung lesions using GLCM homogeneity, contrast, and energy. In our work, GLCM-energy and GLCM-correlation are used to evaluate the texture of synthetic liver lesions. However, GLCM features consider more the distribution of grayscale but ignore the continuity of grayscale and the relationship of grayscale between neighbors. To compensate for this problem, future work may consider using more radiomics features, such as GLSZM, GLRLM, NGTDM, and GLDM(Van Griethuysen *et al* 2017).

On the other hand, the network was trained with 2D images due to hardware limitations and data scarcity. This strategy didn't make full use of the three-dimensional property of CT images. Future studies using 3D image input may improve synthesis quality, particularly in terms of the lesion texture transition along the image thickness direction. In addition, only one constraint parameter of lesions, shape information, is introduced to supervise the synthesis network. In the future, more constraint parameters, such as density, energy, and other radiomics features, can be used to finetune the appearance of synthetic lesions.

## 5. CONCLUSIONS

In this paper, we present a novel lesion synthesis approach to generate labeled training image samples for enhanced lesion segmentation. The synthetic lesions are generated on healthy image slices using automatically generated contours on a new partial convolution generative adversarial network. The segmentation performance is significantly improved after the training dataset is augmented with the synthetic lesions. The approach shows great potential to alleviate the "data paucity" problem in image-based lesion segmentation.

## ACKNOWLEDGEMENTS

Research reported in this publication is supported by the Fundamental Research Funds for the Central Universities (Grant No. WK2030000037), Anhui Provincial-level S&T Megaprojects (Grant No. BJ2030480006).